\documentclass{ws-ijmpd}
\usepackage[super,compress]{cite}
\usepackage{url}
\usepackage{epstopdf}

\begin{document}

\markboth{R. G. Landim}
{Dark energy,  scalar singlet dark matter and the Higgs portal}

%
\catchline{}{}{}{}{}
%

\title{Dark energy,  scalar singlet dark matter and the Higgs portal}

\author{Ricardo G. Landim}

\address{
Instituto de F\'isica, Universidade de S\~ao Paulo\\
 Rua do Mat\~ao,
1371, Butant\~a, CEP 05508-090, S\~ao Paulo, SP, Brazil\label{addr1}\\
 Departamento de F\'isica e Astronomia,\\
Faculdade de Ci\^encias da Universidade do Porto,\\
Rua do Campo Alegre 687, 4169-007, Porto, Portugal\label{addr2}\\
rlandim@if.usp.br}

\maketitle

\begin{history}
\received{Day }
\revised{Day Month Year}
\end{history}

\begin{abstract}One of the simplest extensions of the Standard Model (SM) comprises the inclusion of a massive real  scalar field, neutral under the SM gauge groups, to be a dark matter candidate. The addition of a dimension-six term into the potential of the scalar dark matter enables the appearance of a false vacuum that describes the cosmic acceleration. We show that  the running of the singlet self-interaction and the Higgs portal coupling differs from the standard scalar singlet dark matter model. If we maintain a positive quartic coupling, it is also possible to describe the accelerated expansion of the Universe through a false vacuum with the addition of  a dimension-eight interaction term. In this case,  where the potential remains bounded from below at low energies, the false vacuum decay is highly suppressed.
\end{abstract}


\ccode{PACS numbers: 95.36.+x}


\section{Introduction}\label{sec:level1}

The Standard Model (SM)  is successful to describe the electromagnetic, weak and strong interactions, providing the understanding of several phenomena. However, it is deficient to furnish candidates for the dark sector, which corresponds today to ninety five percent of the energy content of the Universe \cite{Ade:2015xua}. The discovery of the Higgs boson \cite{Aad:2012tfa,Chatrchyan:2012xdj} opened new avenues to investigate the dark sector through interactions with a dark matter (DM) particle. 

One of the simplest extensions of the SM comprises the addition of a massive real scalar field, neutral under the SM gauge group and symmetric under $Z_2$, to play the  role of DM  \cite{Silveira:1985rk,McDonald:1993ex,Burgess:2000yq}. This  scalar singlet $S$ interacts with the SM only through the Higgs portal, $S^2|H|^2$, and has a variety of implications in  different contexts \cite{Bento:2000ah,Bertolami:2007wb,Bento:2001yk,MarchRussell:2008yu,Biswas:2011td,Costa:2014qga,Eichhorn:2014qka,Khan:2014kba,Queiroz:2014yna,Kouvaris:2014uoa,Salvio:2015cja,Bhattacharya:2016qsg,Bertolami:2016ywc,Campbell:2016zbp,Heikinheimo:2016yds,Kainulainen:2016vzv,Nurmi:2015ema,Tenkanen:2016twd,Casas:2017jjg,Arcadi:2017kky,Cosme:2017cxk,Heikinheimo:2017ofk}, including thermal production and annihilation signals \cite{Yaguna:2008hd,Arina:2010rb,Profumo:2010kp}, inflation \cite{Lerner:2009xg,Enqvist:2014zqa,Herranen:2015ima,Kahlhoefer:2015jma,Tenkanen:2016idg}, baryogenesis \cite{Profumo:2007wc,Barger:2008jx,Cline:2012hg} and direct detection through Higgs decay \cite{Mambrini:2011ik,Pospelov:2011yp,Duerr:2015aka,Han:2015dua,Han:2015hda}. Constraints from   several experiments and cosmological observations have been imposed on the scalar singlet DM model. Among a range of studies are the ones using data from  XENON100 and WMAP \cite{Djouadi:2011aa,Cheung:2012xb}, LHC \cite{Djouadi:2012zc,Cline:2013gha,Endo:2014cca}, anti-proton \cite{Goudelis:2009zz,Urbano:2014hda}, LUX \cite{Akerib:2015rjg}, LUX and PandaX \cite{He:2016mls,Escudero:2016gzx}, \textit{Planck} \cite{Ade:2015xua,Cline:2013fm,Slatyer:2015jla} and \textit{Fermi}-LAT \cite{Ackermann:2015zua}. Recently, GAMBIT collaboration used results and likelihoods from  LUX \cite{Akerib:2016vxi}, PandaX \cite{Tan:2016zwf}, SuperCDMS \cite{Agnese:2014aze}, XENON100 \cite{Aprile:2012nq} and  IceCube limits on
DM annihilation to neutrinos \cite{Aartsen:2012kia,Aartsen:2016exj}  to provide
the most stringent constraints to date on the parameter
space of the scalar singlet DM model \cite{Athron:2017kgt}.

The other component of the dark sector, dark energy (DE), is observationally well-described by its simplest candidate, the cosmological constant \cite{Ade:2015xua}. However, the theoretical origin of such a small constant is one of the major concerns in  modern cosmology.
Recently, we have proposed a model of metastable DE \cite{Landim:2016isc} in which the cosmological constant is the difference between a false vacuum state and the true one. Other  models of metastable DE are found in Refs. \cite{Stojkovic:2007dw,Greenwood:2008qp,Abdalla:2012ug,Shafieloo:2016bpk,Stachowski:2016zpq,Szydlowski:2017wlv}. Metastability is also present in the SM, arising from the running of
the Higgs quartic self-coupling. The coupling becomes negative at an energy around $10^{11}$ GeV for the Higgs mass of 125 GeV, indicating that  electroweak (EW) vacuum can decay into a lower energy state, but with a lifetime  longer than the age of the Universe \cite{Cabibbo:1979ay,Sher:1988mj,Sher:1993mf,Isidori:2001bm,EliasMiro:2011aa,Degrassi:2012ry,Alekhin:2012py,Buttazzo:2013uya,Bednyakov:2015sca}. The influence of an additional scalar singlet to the (meta)stability of the EW vacuum was analyzed in Refs. \cite{Lerner:2009xg,Khan:2014kba,Gonderinger:2009jp,Chen:2012faa,Drozd:2013aea,Pruna:2013bma,Belanger:2012zr,Alanne:2014bra,Han:2015hda,Kanemura:2015fra,Robens:2015gla,Robens:2016xkb}.

This paper is divided in two parts. First, we investigate the scalar singlet model with the addition of the potential presented in Ref. \cite{Landim:2016isc}, in the light of the recent GAMBIT collaboration results  \cite{Athron:2017kgt}. This improvement  takes both DM and DE into account in the same model, leading to a cosmological constant that is given in terms of the DM free parameters. Using the renormalization group (RG) equations, we evaluate the running of the scalar DM quartic self-coupling, the running of the coupling with the Higgs boson and the contribution of radiative corrections to the vacuum energy via the Coleman-Weinberg potential \cite{Coleman:1973jx}. In the second part, we consider a positive singlet quartic coupling and show that it is possible to describe the accelerated expansion of the Universe through a false vacuum with the addition of a dimension-eight  interaction term. In this scenario, the potential remains bounded from below at low energies.

This paper is organized as follows. In Sect. \ref{MDE} we expand the previous scalar singlet DM models adding the metastable DE studied in Ref. \cite{Landim:2016isc} and using the recent GAMBIT results. We also present the running of the coupling constants and the contribution due to radiative corrections in the tree-level potential. In Sect. \ref{dim-8} we show that a positive singlet quartic coupling and the addition of a dimension-eight term lead also to a false vacuum.  Sect. \ref{conclu} is reserved for conclusions.

\section{Metastable dark energy and scalar dark matter}\label{MDE}

The extra terms  in the SM Lagrangian due to a massive real scalar field $S$, which have gauge, Lorentz and $Z_2$ symmetries, are 
\begin{equation}\label{Lag}
\mathcal{L}= \mathcal{E}_0+\frac{1}{2}\partial_\mu S \partial^\mu S +\frac{\mu_S^2}{2}S^2+\frac{\lambda_S}{4!} S^4+\frac{\lambda_{hS}}{2}S^2|H|^2+\frac{g'}{M_P^2}S^6\, ,
\end{equation}
where the constant $\mathcal{E}_0$ guarantees a Minkowski true vacuum and $M_P$ is the  Planck mass. In order to have a false vacuum at $S=0$ the quartic self-coupling $\lambda_S$ should be negative, in opposition  to previous scalar singlet DM models, where $\lambda_S\geq 0$. The potential is bounded from below due to the presence of the gravity-induced term $g' M_{P}^{-2}S^6$, which may  parametrizes  a graviton loop contribution \cite{ArkaniHamed:2008ym}. After the spontaneous symmetry breaking the Higg portal leads to an interaction term $h^2S^2$, where $h$ is the physical Higgs boson, and to a singlet mass
\begin{equation}
m_S=\sqrt{\mu_S^2 +\frac{1}{2}\lambda_{hS}v^2}\, ,
\end{equation}
where $v=246$ GeV is the vacuum-expectation-value of the Higgs field.

The difference between the true vacuum and the false one is better seen if we redefine the  coupling of the dimension-six interaction term $g'$:
\begin{equation}\label{VScalar6}
\frac{g'}{M_P^2}S^6=\left[\frac{\lambda_S^2}{2(4!)^2 m_S^2}-\frac{1}{6!}\frac{g}{ M_P^2}\right]S^6\, .
\end{equation}

 The true vacuum, given by $\langle S \rangle_t=\frac{2 \sqrt{6} m_S}{\sqrt{|\lambda_S}|}$, does not affect the Higgs mass  due to  the Higgs portal term $h^2S^2$ because in this model the Universe is in the false vacuum state. The energy of the false vacuum, i.e, at $\langle S \rangle_f=0$, is \cite{Landim:2016isc}
\begin{equation}
\mathcal{E}_0 =\frac{96  m_S^6 g}{5|\lambda_S|^3 M_P^2}\sim 10^{-47} \,  GeV^4\, , 
\end{equation}
and the false vacuum decay is highly suppressed for $m_S\geq 10^{-12}$ GeV.

In order to investigate the possibility of this model to describe both DM and DE we take the best-fit values given by the GAMBIT collaboration  \cite{Athron:2017kgt}, for which the scalar field $S$ constitutes the entire observed relic density of DM ($\Omega_Sh^2\sim\Omega_{DM}h^2$). Using the \textit{Planck} results $\Omega_{DM}h^2= 0.119 $ \cite{Ade:2015xua} and perturbative couplings, the scalar field mass  is $m_S= 62.27$ GeV and the  Higgs portal coupling is $\lambda_{hS}=2.9 \times 10^{-4}$. With this value of scalar mass and with a singlet quartic self-interaction of around $|\lambda_S|\sim 1$, for instance, the observed vacuum energy is reached for $g\sim 10^{-23}$. Smaller values of $\lambda_S$ leads to smaller values of $g$.

\subsection{ Radiative corrections}

\subsubsection{Renormalization Group Equations}

Considering the renormalizable term in the Lagrangian, a negative $S$ self-interaction coupling ($\lambda_S<0$) changes the evolution of $\lambda_S$ itself  and the Higgs portal coupling $\lambda_{hS}$ with a  RG scale $\mu$. The beta-function of the Higgs self-interaction $\lambda$ has an additional term $\lambda_{hS}^2/2$  at one-loop due to  the singlet \cite{Haba:2013lga}, however due to the smallness of it,  the running of the Higgs self-coupling is not affected \cite{Athron:2017kgt}. On the other hand, the beta-functions $\beta_i\equiv (4\pi)^2 \frac{d\lambda_i}{d\ln \mu}$ for $\lambda_S$  and $\lambda_{hS}$ are \cite{Haba:2013lga}
\begin{equation}
\beta_S=3\lambda_S^2+12\lambda_{hS}^2\, ,
\end{equation}
\begin{equation}
\beta_{hS}=\lambda_{hS}\left(4\lambda_{hS}+12 \lambda+6y^2-\frac{3}{2}(g_1^2+3g_2^2)+\lambda_S\right)\, ,
\end{equation}
where the RG scale $\mu$ runs from EW to Planck scale, $y$ is the Yukawa coupling, $\lambda$ is the Higgs quartic self-interaction and $g_1$ and $g_2$ are the SM couplings. 

\begin{figure}[ht]
\includegraphics[scale=0.65]{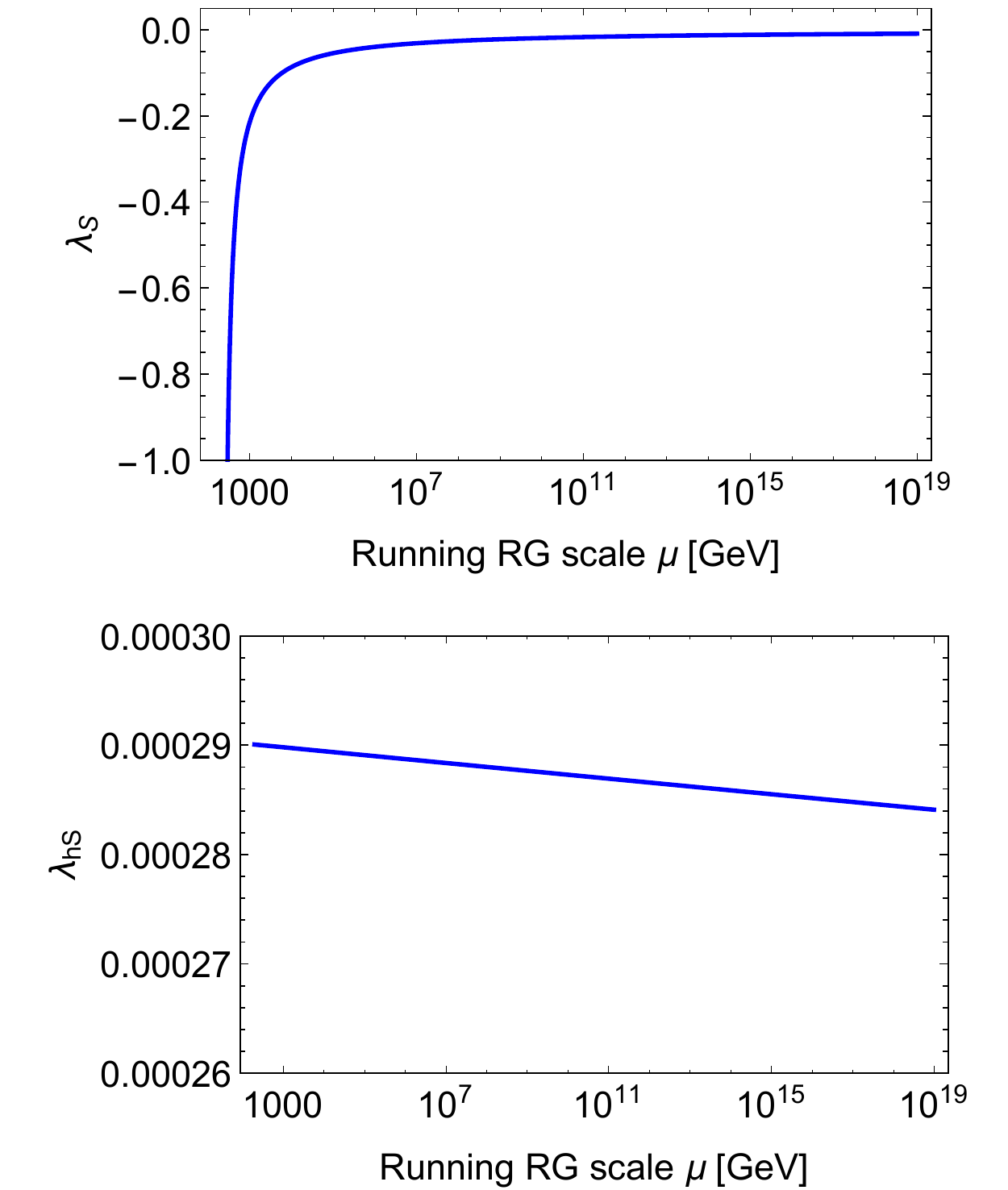}
\caption{RG evolution of the singlet self-interaction $\lambda_S$ (top) and the Higgs portal coupling $\lambda_{hS}$ (bottom) in the $\overline{MS}$ scheme. For illustrative purposes, the couplings used were $g_1=0.4$, $g_2=0.7$, $\lambda=0.1$, $\lambda_S=-1$ and $y_t=0.6$.}\label{fig:running}
\end{figure}

The running of the singlet self-interaction $\lambda_S$ and the running of the Higgs portal coupling  $\lambda_{hS}$ are shown in Fig. \ref{fig:running}. The quartic coupling $\lambda_S$ increases with the RG scale but remains always negative. For $\lambda_S>0$  the Higgs portal coupling  $\lambda_{hS}$   increases with the RG scale \cite{Haba:2013lga}, but in our case $\lambda_S<0$ and $\lambda_{hS}$ decreases slightly.

\subsubsection{Coleman-Weinberg Potential}
The tree-level potential in Eq. (\ref{Lag})  receives radiative corrections at one-loop via the Coleman-Weinberg potential \cite{Coleman:1973jx,Ford:1992mv}, becoming $V=V^{(0)}+V^{(1)}$, where $V^{(1)}$ has the  following form for a generic scalar field $\chi$
\begin{equation}
V^{(1)}(\chi)=\sum_i{\frac{n_i}{64\pi^2}M_i(\chi)^4}\left[\ln\left(\frac{M_i^2(\chi)}{\mu^2}\right)-c_i\right]\, ,
\end{equation}
 where $``i''$ runs over all degrees of freedom in the model, $n_i$ is the number of degrees of freedom,  $M_i^2(\chi)=\kappa_i\chi^2-\kappa'_i$, $\kappa_i$, $\kappa'_i$ and $c_i$ are constants. Here we considered a fixed RG scale $\mu$. The correction for the Higgs direction, $\chi=h$, in the SM$+S$ model is given in Ref. \cite{Lerner:2009xg} and it is unchanged for the case $\lambda_S<0$. On the other hand, in the $S$ direction the one-loop potential does change, due to the following contribution
\begin{equation}
V^{(1)}(S)=\frac{1}{64\pi^2}M_S(S)^4\left[\ln\left(\frac{M_S^2(S)}{\mu^2}\right)-\frac{3}{2}\right]\, ,
\end{equation}
where $M_S^2(S)=\mu_S^2+\frac{\lambda_S}{2}S^2$. At the false vacuum, $S=0$, the one-loop correction in the $S$ direction gives a contribution  to the vacuum energy of $V^{(1)}(0)\sim -10^6$ GeV$^4$, for $\mu\sim 300$ GeV. This result is insensitive to different RG scales  and  it overlaps  completely the observed value of the vacuum energy ($10^{-47}$ GeV$^4$). However, the tree-level potential contains a non-renormalizable term ($g' M_P^{-2}S^6$), needed to stabilize the potential but which spoils a traditional attempt to include radiative corrections to the proposed scalar singlet DM model. Therefore, although there is a huge contribution to the vacuum energy given by radiative corrections that in principle could be canceled by extra terms, the inclusion of the dimension-six term leads to a consistent model for dark energy at the tree-level.

\section{Metastable dark energy with a dimension-eight term}\label{dim-8}

\begin{figure}%
\includegraphics[scale=0.65]{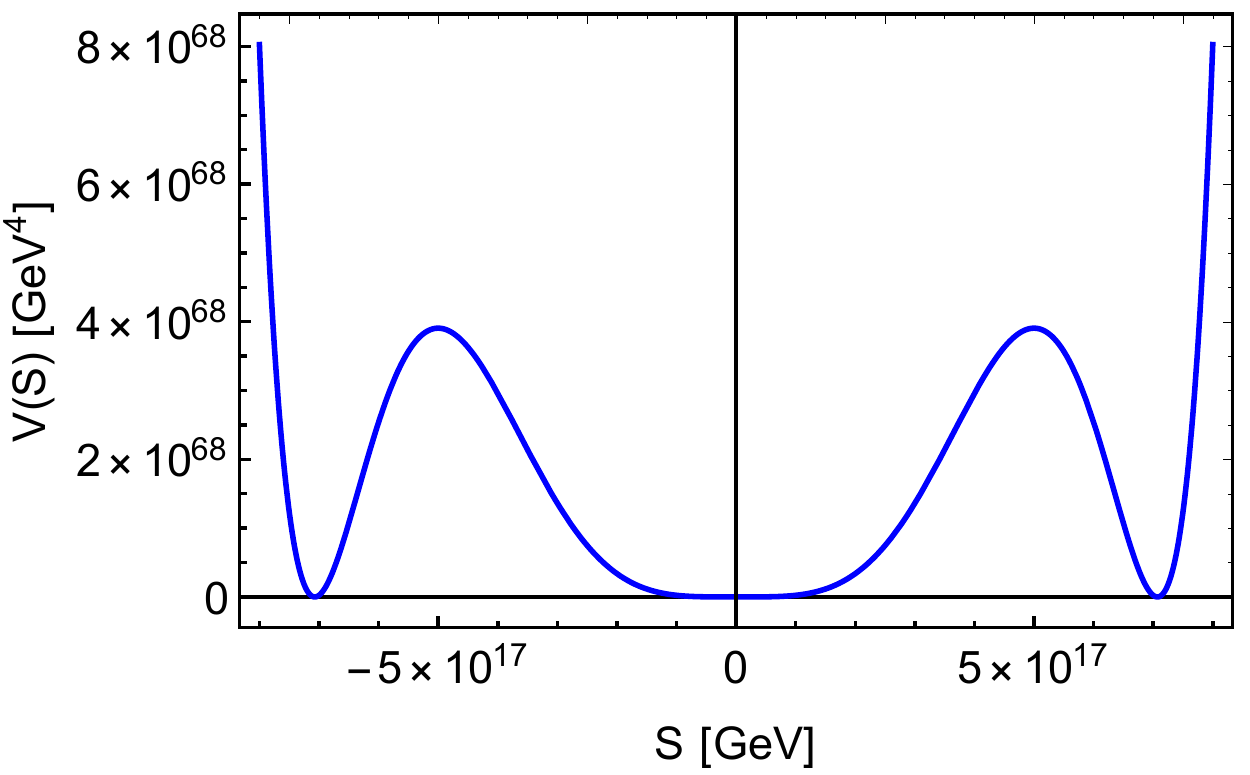}%
\caption{Scalar potential, Eq. (\ref{eq:pot8}), with arbitrary values for the free parameters. The difference between the false vacuum, at $S=0$, and the true one, at $S=\pm 7\times 10^{17}$ GeV, is the observed vacuum energy ($10^{-47}$ GeV$^4$, not shown). Notice that the mass term is sub-dominant.}%
\label{pot8}%
\end{figure}

Instead of considering a negative singlet quartic coupling we assume now a stable potential at low energies, thus recovering the results in the literature for the case $\lambda_S>0$. Even in this scenario is still possible to have a false vacuum if we add a dimension-eight  term to the potential showed in the last section. The potential for the singlet $S$ after the symmetry breaking becomes
\begin{equation}
V(S)=\mathcal{E}_0+\frac{m_S^2}{2}S^2+\frac{\lambda_S}{4!} S^4+\frac{g_6}{\Lambda^2}S^6+\frac{g'_8}{\Lambda^4}S^8\, ,
\label{eq:pot8}
\end{equation}
where $\Lambda $ is a cutoff scale and $g_6<0$. The potential is depicted in Fig. \ref{pot8}. The mass term can be neglected in the following calculations because the true vacuum is at large values of the field $S$, as can be seen in the figure and below. 

The dimension-eight term can be rewritten as
\begin{equation}\label{VScalar8}
\frac{g'_8}{\Lambda^4}S^8=\left[\frac{6g_6^2}{\Lambda^4\lambda_S}-\frac{g_8}{ \Lambda^4}\right]S^8\, ,
\end{equation}
in such a way that the last term in the right-hand side causes a deviation from the Minkowski vacuum. The true vacuum is given by $\langle S\rangle_t=\sqrt{\frac{\lambda_S}{3|g_6|}}\frac{\Lambda}{2}$ and the energy of the false vacuum is
\begin{equation}
\mathcal{E}_0 =\frac{g_8 \lambda_S^4\Lambda^4}{20736 g_6^4}\sim 10^{-47} \,  GeV^4\, . 
\end{equation}

In order to the false vacuum to be long-lived and describe the cosmic acceleration, the decay rate should be highly suppressed. The decay rate per unit of volume is given by  \cite{Coleman:1977py}
\begin{equation}
\frac{\Gamma}{V}\sim \exp(-S_E)\, ,
\label{eq:decayrate}
\end{equation}
where $S_E$ is the Euclidean action for the singlet. Using the thin-wall approximation, the Euclidean action for a stationary bubble radius is
\begin{equation}
S_E\simeq \frac{27\pi^2}{2\epsilon^3}\left[\sqrt{2}\int_{\langle S\rangle_f}^{\langle S\rangle_t}\,d S{ \sqrt{V(S)}}\right]^4\, ,
\label{eq:SthinA}
\end{equation}
where $\epsilon $ is the observed vacuum energy. Using the potential (\ref{eq:pot8}), the Euclidean action yields
\begin{equation}
S_E\sim \frac{10^{-10}\lambda_S^8\Lambda^{12}}{g^6_6\epsilon^3}\, .
\label{eq:SthinAresult}
\end{equation}
The action is very large even for a cutoff of 10 TeV, thus the decay is highly suppressed. 

Considering the gravitational effect in the computation of the decay rate, we have a new action  $\bar{S}_E$ given in terms of the old one $S_E$  \cite{Coleman:1980aw}
\begin{equation}
\bar{S}_E=\frac{S_E}{\left(1+\left(\frac{R}{2\Delta}\right)^2\right)^2}\, ,
\label{Sbar}
\end{equation}
where $R\simeq 3\sqrt{2}\epsilon^{-1}\int_{\langle S\rangle_f}^{\langle S\rangle_t}\,d S{ \sqrt{V(S)}}$ is the bubble radius and $\Delta=\frac{\sqrt{3}M_{P}}{\sqrt{\epsilon}}$ is the value of the bubble radius when it is equal to the   Schwarzschild radius. The gravitational correction given by $(R/(2\Delta))^2\sim 10^{-6}\lambda_S^4\Lambda^4 \epsilon^{-1}g_6^{-3}$ is not negligible  and should be taken into account. For illustrative purposes, using conservative values $\lambda_S\sim 1$, $\Lambda\sim 10$ TeV and $g_6 \sim 1$, the actions have very large values,  $S_E\sim 10^{191}$ and  $\bar{S}_E\sim 10^{69}$. Larger cutoff scales lead to larger values for both actions. Therefore,  the decay rate is still highly suppressed and the false vacuum is long-lived.

\section{Conclusions}\label{conclu}

In this paper we have included a dimension-six term in the usual scalar singlet DM model. By considering a negative singlet quartic self-interaction coupling it is possible to describe the accelerated expansion of the Universe through a false vacuum using the recent GAMBIT collaboration results. The running of the singlet self-coupling and the Higgs portal coupling have been evaluated and have shown different behaviors from the ones of positive $\lambda_S$. In the case of a positive   quartic coupling it is possible to describe the cosmic acceleration through the addition of a dimension-eight term. In this scenario, where  the potential remains bounded from below at low energies, the false vacuum plays the role of the cosmological constant since its decay is highly suppressed.

\section*{Acknowledgments}
This work is supported by CNPq (Conselho Nacional de Desenvolvimento Cient\'ifico e Tecnol\'ogico, grant 150254/2017-2). RL thanks Orfeu Bertolami and the Departamento de F\'isica e Astronomia, Faculdade de Ci\^encias, Universidade do Porto for hosting him while the work was in progress.

\bibliographystyle{ws-ijmpd}
\bibliography{references}

\end{document}